\documentclass[twoside]{article}
\usepackage{fleqn,espcrc2}

\usepackage{graphicx}
\usepackage[figuresright]{rotating}

\newcommand{\AmS}{{\protect\the\textfont2
  A\kern-.1667em\lower.5ex\hbox{M}\kern-.125emS}}

\title{$B_{s,d}\rightarrow\gamma\gamma$ decay in the model with one
  universal extra dimension}

\author{        G.  Devidze\address[IHEPI]{Institute of High Energy Physics and Informatization,\\ 
        9 University St., 0186 Tbilisi, Georgia}
        \thanks{Present address: Institut f\"ur Kernphysik,  
                Forschungszentrum J\"ulich, 52425 J\"ulich, Germany},
                A. Liparteliani\addressmark[IHEPI]
        \thanks{Present address: Institut f\"ur Kernphysik, 
                Forschungszentrum J\"ulich, 52425 J\"ulich, Germany},
        and
        Ulf-G. Mei$\ss$ner\address{Universit\"at Bonn, Helmholtz-Institut f\"ur Strahlen- und Kernphysik         (Theorie)\\
        D-53115 Bonn, Germany}$^,$\address{
        Institut f\"ur Kernphysik (Theorie), Forschungszentrum J\"ulich, 52425 J\"ulich, Germany}
}

\begin{document}

\begin{abstract}

We estimate the beyond the Standard 
Model (SM) contribution to the $B_{s,d}\rightarrow\gamma\gamma$ 
 double radiative decay in the 
framework of the model with one universal extra dimension. 
This contribution gives a $\sim 3\,(6)\%$ enhancement of the branching ratio 
calculated in the SM for $B_{s\,(d)} \to \gamma\gamma$.

\vspace{1pc}
\end{abstract}

\maketitle

\section{Introduction}

It is known that in the Standard Model (SM) the double radiative 
decays of the $B_{s,d}$  mesons, $B_{s,d}\rightarrow\gamma\gamma$,  
first arise at the one loop level with the exchange of up-quarks and 
W-bosons in the 
loops~\cite{Lin:1990kw,Simma:1990nr,Herrlich:1991bq,Devidze:1996uw,Devidze:1996um}. 
The branching ratios for the above decays are of the 
order of $\sim 10^{-7} \,(10^{-9})$.

On the other hand there is the possibility to enhance the above mentioned decays
in  extended versions of the SM. 
In the  papers \cite{Devidze:1998cf,Devidze:2000he} it was shown 
that in supersymmetric versions of the SM one could reach a branching 
ratio as large as ${\rm Br}(B_s\rightarrow\gamma\gamma)\sim 10^{-6}$
 depending on the SUSY parameters. This enhancement 
was achieved mainly due to the exchange of charged scalar Higgs particles 
within the loop. There exists an analogous possibility in other exotic 
models as well for the scalar particle exchange inside the loop, which 
could potentially enhance this process. For example, the Appelquist, 
Chang and Dobresku (ACD) model with only one universal 
extra dimension \cite{Appelquist:2000nn} 
presents us with such an  opportunity. One should note that in the above
approach towers of charged Higgs particles arise as real objects with 
certain masses, not as fictitious (ghost) fields.

In this letter we aim to calculate the contributions from these
real scalars to the $B_{s,d}\rightarrow\gamma\gamma$
decay. The article is organized as follows: 
in the section~2 some useful information about the  ACD model, necessary for 
the calculations, is provided. Section~3 is devoted to the calculation 
of the pertinent amplitudes. In  section~4, numerical estimates of 
the branching ratios are discussed.

\section{Useful information on the structure of the ACD-model}

In the Universal Extra Dimension (UED) scenarios all the fields presented in 
the SM live in extra dimensions, i.e. they are functions of all
space-time coordinates. For bosonic fields one simply replaces all 
derivatives and fields in the SM lagrangian by their 5-dimensional 
counterparts. These are the $U(1)_Y$-and $SU(2)_L$-gauge fields as 
well as the $SU(3)_C$-gauge 
fields from the QCD -sector. The Higgs doublet is chosen to be even 
under $P_5$ ($P_5$ is a parity operator in the five dimensional space) and 
possesses a zero mode. Note that all zero modes remain massless 
before the Higgs mechanism is applied. In addition we should note 
that as a result of the action of the parity operator the fields receive 
additional masses $\sim n/R$ after dimensional reduction and transition to the 
four dimensional Lagrangians.

In the five dimensional ACD model the same  procedure for gauge 
fixing is possible as in the 
models in which fermions are localized on 
the 4-dimensional subspace. With the gauge fixed, one can diagonalize 
the kinetic terms of the bosons and finally derive expressions for the 
propagators. Compared to the SM, there are additional Kaluza-Klein
(KK) mass terms. As they are common to all fields, their contributions 
to the gauge boson mass matrix is proportional to the unity matrix. As 
a consequence, the electroweak angle remains the same for all KK-modes 
and is the ordinary Weinberg angle  $\theta_W$. Because of the 
KK-contribution to the mass matrix, charged and neutral Higgs
components with $n\not=0$ ($n$ being the 
number of the KK-mode) no longer play the role of Goldstone bosons. 
Instead, they mix with $W_5^{\pm}$  and $Z_5$ to form, in addition to 
the Goldstone modes $G^0_{(n)}$  and $G^{\pm}_{(n)}$,  three additional 
physical states $a^0_{(n)}$  and $a^{\pm}_{(n)}$ . It is precisely the role of 
these additional charged physical states to  double 
radiative neutral  $B$-meson decays that is studied in this paper. 

The Lagrangian responsible for the interaction of charged scalar KK
towers   $a^{*}_{(n)} $ with the ordinary down quarks reads
\begin{eqnarray}
{\mathcal L}=\frac{g_2}{\sqrt{M_{(n)}}}\bar{Q}_{i(n)}
(C_L^{(1)}P_L+C_L^{(1)})a^*_{(n)}d_j+\nonumber \\
  \frac{g_2}{\sqrt{M_{(n)}}}\bar{U}_{i(n)}
(C_L^{(2)}P_L+C_L^{(2)})a^*_{(n)}d_j~,        
\end{eqnarray}
%
utilizing the following notations \cite{Buras:2002ej}:
\begin{eqnarray}
C_L^{(1)}&=&-m_3^{(i)}V_{ij}, \hspace{9mm}
C_L^{(2)}=m_4^{(i)}V_{ij},  \nonumber\\
C_R^{(1)}&=&Mm_3^{(i,j)}V_{ij}, \hspace{2mm
C_R^{(2)}=-M_4^{(i,j)}V_{ij}}, \nonumber\\
M^2_{W(n)}&=&m^2(a^*_{(n)})=M^2_W+\frac{n^2}{R^2},        
\end{eqnarray}
where $V_{ij}$ are elements of the CKM matrix. The mass parameters 
in Eq.(2) are defined as
\begin{eqnarray}
m_3^{(i)}&=&-M_Wc_{i(n)}+\frac{n}{R}\frac{m_i}{M_W}s_{i(n)},\nonumber\\
m_4^{(i)}&=&M_Ws_{i(n)}+\frac{n}{R}\frac{m_i}{M_W}c_{i(n)},\nonumber\\
M_3^{(i,j)}&=&\frac{n}{R}\frac{m_j}{M_W}c_{i(n)},\nonumber\\
M_4^{(i,j)}&=&\frac{n}{R}\frac{m_j}{M_W}s_{i(n)}.     
\end{eqnarray}
Here, $M_W$ and the masses of up (down)-quarks $m_i \, (m_j)$ in the 
right-hand-side of Eq.(3) are zero mode masses and the $c_{i(n)}$, $s_{i(n)}$ 
stand for the $\cos$ and $\sin$ of the fermions mixing angles, respectively,
\begin{equation}
\tan 2\alpha_{f(n)}=\frac{m_f}{n/R},\hspace{3mm} n\geq1~.   
\end{equation}
The masses for the fermions are calculated as
\begin{equation} 
m_{f(n)}=\sqrt{\frac{n^2}{R^2}+m^2_f} . 
\end{equation}
In the phenomenological applications we use the restriction 
$n/R \geq 250\,\mbox{GeV}$ and hence we assume that all the fermionic
mixing angles except $\alpha_{t(n)}$ are equal zero.

\section{Structure of  $B_{s,d}\rightarrow\gamma\gamma$  in the ACD
  model with one extra dimension}

The Feynman graphs, describing the contributions of scalar physical
states to process under consideration, are shown in Fig.1. 
\begin{figure}[htb]
\centering 
\includegraphics[width=60mm]{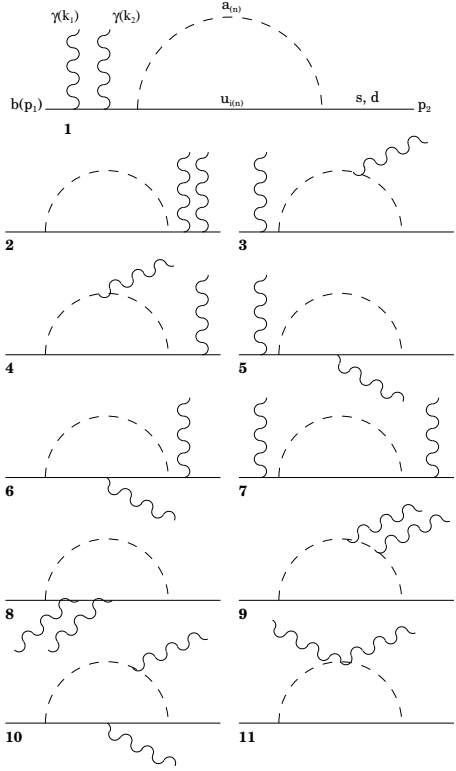}
\caption{Double radiative $B$-meson decay
  $B_{s,d}\rightarrow\gamma\gamma$ 
in the theory with only one extra universal dimension (the dashed lines in
  the loops correspond to the charged KK towers $a^*_{(n)}$, while the
  solid lines in the loops are for the up-quark KK towers).}
\label{fig:toosmall}
\end{figure}

The amplitude for the decay 
$B_{s,d}\rightarrow\gamma\gamma$ has the form
\begin{eqnarray}
T(B\rightarrow\gamma\gamma) &=&\epsilon_1^{\mu}(k_1)\epsilon_2^{\nu}(k_2)
\nonumber \\
&& \quad \times[Ag_{\mu\nu} + iB\epsilon_{\mu\nu\alpha\beta}k_1^{\alpha}k_2^{\beta}]. 
\end{eqnarray}
This equation is correct after gauge fixing for the final photons 
which we have chosen as
\begin{equation}
\epsilon_1 \cdot k_1=\epsilon_2 \cdot k_2=\epsilon_1 \cdot k_2
=\epsilon_2 \cdot k_2=0, 
\end{equation}
where $\epsilon_1$ and $\epsilon_2$ are photon polarization vectors, respectively.
The condition Eq.(7) together
with  energy-momentum conservation leads to
\begin{equation}
\epsilon_i \cdot P=\epsilon_i \cdot p_1=\epsilon_i\cdot p_2=0,  
\end{equation}
where
\begin{equation}
P=k_1+k_2 \hspace{3mm} \mbox{and}\hspace{3mm} 
p_1=p_2+k_1+k_2.     
\end{equation}                    
Let us write down some useful kinematical relations which are results 
of Eqs.(7,8) as well:
\begin{eqnarray}
P\cdot p_1 &=&m_bM_B,\quad  P\cdot p_2 =-m_{s(d)}M_B,  \nonumber\\
P\cdot k_1 &=& P\cdot k_2 = k_1\cdot k_2 =\frac12M^2_B, \nonumber\\
p_1 \cdot p_2 &=&-m_bm_{s(d)},\nonumber \\
p_1\cdot k_1 &=& p_1 \cdot k_2 = \frac12m_bM_B, \nonumber \\
p_2 \cdot k_1 &=& p_2 \cdot k_2 =-\frac12m_{s(d)}M_B. 
\end{eqnarray}

The total contributions into $CP$-even ($A$) and $CP$-odd ($B$)
 amplitudes from 
Eq.(6) are calculated as sums of the  appropriate contributions
of the diagrams in Fig.1 corresponding to a tower of scalar particle
contributions in the ACD model with only one extra dimension.
Let us note that we used the following formula for the hadronic 
matrix elements:
\begin{equation}
\langle 0 | {\bar s\,(\bar d)} \gamma_{\mu}\gamma_5 b | B(P)\rangle=
-if_BP_{\mu}.   
\end{equation}

Apart from one particle reducible (1PR) diagrams, one particle
irreducible (1PI) ones contribute to the amplitudes, and hence, 
to their $CP$-even ($A$) and  $CP$-odd ($B$) parts. 
We should note that each of 
the 1PI contributions is finite. Let us discuss these contributions 
in more details. In the SM only one 1PI diagram (one with the $W$-boson 
exchange in the loop, when both photons are emitted by virtual
up-quarks) gives the contribution of the order of
$\sim 1/M^2_W$ . In the Ref.\cite{Gaillard:1974hs}
it was observed that diagrams with  light quark exchange contribute 
as $\sim1/M^2_W$, 
while diagrams containing the heavy quarks are of order of $\sim1/M^4_W$. 
In the ACD  model the contributions of such diagrams are of the order of
$\sim1/M^4_W$ because 
the estimate for all KK-tower masses, including the ones exchanged in 
the loops, in our case are $M\geq 250\,$GeV. 
Likewise discussions show that all
the 1PI diagrams existing in the ACD model also are of order $\sim1/M^4_W$. 
Thus, the leading 1PI diagrams are negligible and we do not consider them.

\section{Branching ratio for the $B\rightarrow\gamma\gamma$  decay}

The total contributions to the $B\rightarrow\gamma\gamma$  decay amplitudes are:
\begin{eqnarray}
A&=&b\frac{m_b^3}{m_{s,(d)}}\Bigl{\{}
\frac{n}{RM_W}m^{(i)}_3m_{i(n)}c_{i(n)}f_1(x_i)+ \nonumber \\
&& \bigl[ (m_3^{(i)})^2
-\frac{n^2}{R^2M^2_W}m_bm_{s(d)}c^2_{i(n)}\bigr]\frac12 f_2(x_i)
\Bigr{\}}, \nonumber \\
B&=&2b\frac{m_b}{m_{s,(d)}}\Bigl{\{}
\frac{n}{RM_W}m^{(i)}_3m_{i(n)}c_{i(n)}f_1(x_i)+ \nonumber \\
&&\bigl[ (m_3^{(i)})^2
+\frac{n^2}{R^2M^2_W}m_bm_{s(d)}c^2_{i(n)}\bigr]\frac12 f_2(x_i)
\Bigr{\}}, \nonumber\\ &&
\end{eqnarray}                        
where
\begin{eqnarray}
b&=&\frac14\frac{i}{(4\pi)^2}e^2g^2_2f_B\frac{Q_d}{M^2_{W(n)}}
\frac{V^*_{is(d)}V_{ib}}{m^2(a^*_{(n)})}, \nonumber\\
f_1(x)&=&\frac{-5x^2+8x-3+2(3x-2)\ln x}{6(1-x)^3}, \nonumber\\
f_2(x)&=&\frac{-2x^3-3x^2+6x-1+6x^2\ln x}{6(1-x)^4},\nonumber\\
x_i&=&\frac{m^2(u_{i(n)})}{m^2(a^*_{(n)})}~.
\end{eqnarray}
As it is obvious from Fig.1, the correct calculation assumes 
the inclusion of the crossed diagrams (not shown on fig.1).
In the kinematics we use, cf. Eqs.(7-10), this leads to  a factor 2
for all amplitudes, except for the one given by diagram 11.
 However, diagram 11 belongs to the class of 1PI diagrams. As it
was stated above, one particle irreducible diagrams does not give
leading contributions into process and therefore their contributions
($\sim1/M^4_W$) are negligible comparing with that of the 1PR diagrams.

        On the other hand, using the unitarity feature of the 
Kobayashi-Cabibbo-Maskawa matrix, the amplitude for double radiative
$B$-meson decay can be rewritten as:
\begin{equation}
T=\sum_{i=u,c,t}\lambda_iT_i=\lambda_t\Bigl\{T_t-T_c+
\frac{\lambda_u}{\lambda_t}(T_u-T_c)\Bigr\}.  
\end{equation}

Let us note that we restricted ourselves by calculating the leading 
order terms of $\sim1/M^2_W$ from the up-quark KK-towers. 
In this approximation it 
turns out that the $u_{(n)}$  and the $c_{(n)}$ towers have equal
contributions. 
Therefore, the expressions for the amplitudes have a simpler form than before:
\begin{eqnarray}
A=\lambda_t(A_{t(n)}-A_{c(n)})~,\nonumber\\
B=\lambda_t(B_{t(n)}-B_{c(n)})~.    
\end{eqnarray}
Furthermore, it is easy to obtain from Eq.(6) the expression 
for the $B\rightarrow\gamma\gamma$ decay partial width:
\begin{equation}
\Gamma(B\rightarrow\gamma\gamma)=\frac{1}{32\pi M_B}
\Bigl[ 4 | A |^2+\frac12M^4_B |B|^2   \Bigr].   
\end{equation}
Now we are in the position to compare the ACD contribution to the 
decay with that of the SM. Namely, let us consider the ratio:
\begin{eqnarray}
&& \!\!\!\!\!\!\!\! \!\!\!\!\!\!\!\!
\frac{\Gamma(B_{s(d)}\rightarrow\gamma\gamma)_{\rm ACD}}
{\Gamma(B_{s(d)}\rightarrow\gamma\gamma)_{\rm SM}}=
\frac{24n^2M^6_W}{Q^2_dR^2M^4_{W(n)}m^4(a^*_{(n)})} \nonumber\\
&\times& \Bigl{\{}\frac{m^{(i)}_3m_{i(n)}}{M^2_W}c_{t(n)}f(x_{t(n)})
+ \frac{n}{RM_W}f(x_{c(n)})\Bigr{\}}^2 \nonumber\\
&\Bigl/& 
\Bigl{\{}
4\bigl( C(x_t)+\frac{23}{3} \bigr)^2 
+ 2\bigl(C(x_t)+\frac{23}{3} \nonumber\\
&+&
16\frac{m_{s(d)}}{m_b}\bigr)^2 \Bigr{\}} 
\end{eqnarray}
where
\begin{eqnarray}
C(x)&=&\frac{22x^3-153x^2+159x-46}{6(1-x)^3} \nonumber\\
&+&\frac{3(2-3x)}{(1-x)^4}\ln x, \qquad x_t=\frac{m^2_t}{M^2_W}. 
\end{eqnarray}
Rough numerical estimates of Eq.(17) show that in case of  
$B_s$-meson decay we can get a difference from SM-result as much 
as $\sim 3\%$.  The UED contribution to the $B_{s}\rightarrow\gamma\gamma$ 
is $3\%$ of the SM estimate and increases the overal contribution (SM+UED) 
by $3\%$.
The same difference for the case of 
$B_d\rightarrow\gamma\gamma$ is $\sim 6\%$. 

The theoretical estimates of double radiative $B$-decays in the
framework of the Standard Model, 
${\rm Br}(B_s\rightarrow\gamma\gamma)\sim 10^{-7}$
 and  ${\rm Br}(B_d\rightarrow\gamma\gamma)\sim 10^{-9}$
along with the upper 
experimental limits [11] allows us to hope that in the not to far
future these decays will be observed (say, by the BaBar or BELLE 
collaborations or at the CERN B-physics facility).
We thus hope that not too much  time will pass until the 
differences of $\sim 3\% (6\%)$  will be accessible for 
experimental analysis.

\bigskip

{\bf Acknowledgements}
\vspace{0.5pc}

 The research described in this publication was made possible in part
 by Award No. GEP1-33-25-TB-02 of the Georgian Research and
 Development Foundation (GRDF) and the U.S. Civilian
 Research \& Development Foundation for the Independent States of the
 Former Soviet Union (CRDF). Part of this work have been done due to
 support of Deutscher Akademischer Austauchdienst (DAAD).
  G.D. and A.L. wish to express their great
 attitude to  H. Str\"oher, N.N.~Nikolaev,  C.~Hanhart,
 A.~Kacharava, A.~Wirzba for valuable discussions and help during our 
 stay at IKP-FZJ as well as for excellent scientific environment and warm
 hospitality they provided for us. This research is part of the EU 
Integrated Infrastructure Initiative Hadron Physics Project
under contract number RII3-CT-2004-506078.

\bigskip



\begin{thebibliography}{9}




\bibitem{Lin:1990kw}
G.~L.~Lin, J.~Liu and Y.~P.~Yao,
Phys.\ Rev.\ D {\bf 42} (1990) 2314.
\bibitem{Simma:1990nr}
H.~Simma and D.~Wyler,
Nucl.\ Phys.\ B {\bf 344} (1990) 283.
\bibitem{Herrlich:1991bq}
S.~Herrlich and J.~Kalinowski,
Nucl.\ Phys.\ B {\bf 381} (1992) 501.
\bibitem{Devidze:1996uw}
G.~G.~Devidze, G.~R.~Dzhibuti and A.~G.~Liparteliani,
Nucl.\ Phys.\ B {\bf 468} (1996) 241.
\bibitem{Devidze:1996um}
G.~G.~Devidze, G.~R.~Dzhibuti and A.~G.~Liparteliani,
Phys.\ Atom.\ Nucl.\  {\bf 59} (1996) 1948
[Yad.\ Fiz.\  {\bf 59} (1996) 2021].

\bibitem{Devidze:1998cf}
G.~G.~Devidze and G.~R.~Dzhibuti,
Phys.\ Lett.\ B {\bf 429} (1998) 48.


\bibitem{Devidze:2000he}
G.~G.~Devidze and G.~R.~Dzhibuti,
Phys.\ Atom.\ Nucl.\  {\bf 63} (2000) 310
[Yad.\ Fiz.\  {\bf 63} (2000) 373].

\pagebreak

\bibitem{Appelquist:2000nn}
T.~Appelquist, H.~C.~Cheng and B.~A.~Dobrescu,
Phys.\ Rev.\ D {\bf 64} (2001) 035002
[arXiv:hep-ph/0012100].

\bibitem{Buras:2002ej}
A.~J.~Buras, M.~Spranger and A.~Weiler,
Nucl.\ Phys.\ B {\bf 660} (2003) 225
[arXiv:hep-ph/0212143].

\bibitem{Gaillard:1974hs}
M.~K.~Gaillard and B.~W.~Lee,
Phys.\ Rev.\ D {\bf 10} (1974) 897.

\bibitem{Data} Particle Data Group, http://pdg.lbl.gov.

\end{thebibliography}
\end{document}